# Effective properties of acoustic waves in a poroelastic medium containing spherical cavities randomly distributed within the Rayleigh limit

Dossou GNADJRO(a,b), Amah D'ALMEIDA(a), Hervé FRANKLIN(b)

a. Département de Mathématiques, Université de Lomé, Togo + dgnadjro@gmail.com, dal _me@yahoo.fr
b. Laboratoire Ondes et Milieux Complexes (LOMC) UMR CNRS 6294 Université Le Havre 76600  Le Havre
+ herve.franklin@univ-lehavre.fr

**Abstract**: The propagation of acoustic waves in a poro-elastic medium of infinite extension containing spherical cavities randomly distributed is investigated. The scattering coefficients are computed in the low frequency limit using the sealed pore boundary conditions. Also in this limit, using explicit Waterman-Truell (WT) formulas and a generalization of the Linton-Martin (LM) formula to poro-elastic medium, expressions of the wave numbers of the coherent waves are proposed. Expressions of the effective mass densities and moduli are derived from the WT formulas. The effective properties of the coherent wave in an elastic medium are obtained as a limiting case when the porosity tends towards zero.

**Highlights :** multiple scattering, Rayleigh limit, poro-elastic medium, Effective properties.

## 1  Introduction

The study of multiple scattering of acoustical waves by obstacles has numerous applications in material sciences and in industry and considerable work has been done on the subject. In scattering problems, the significant parameter is $ka$ where $a$ is a characteristic length of the scatter and $k$ the wave number. The Rayleigh limit refers to cases where $ka \ll 1$. The medium supporting the wave can be a fluid, a solid or a porous medium. Many researchers have studied wave scattering by spherical inclusions suspended in a liquid [1][2][7][6][14][15][20] or by spherical inclusions in an elastic or poro-elastic medium [4][11][16]. According to Biot's theory [3] three kinds of waves, a fast (1) and a slow (2) compressional waves and a transverse (t) one, can propagate in a porous medium. An encounter with an obstacle causes the scattering of any of them in the three kinds of waves by mode conversion. The multiple scattering model of Waterman and Truell [6], which is an improvement of that of Foldy [1] for spherical inclusions in a fluid, leads to an expression of the effective wave number for spheres randomly distributed in a fluid medium with a low density $n_0$ (the number of spheres per unit volume). This formula is still valid in a porous medium for each kind of waves considered by neglecting the conversion phenomena [17]. Lloyd and Berry(LB)[7] in 1967 extended the Waterman and Truell(WT) model to spherical inclusions of finite size by introducing the concept of radius of exclusion which prohibits two inclusions from interpenetration. This made it possible to correct the second order term in concentration by an integral taking into account all the angles of scattering. In 2006, Linton and Martin [14] proposed an equivalent writing of the effective wave number of LB by transforming the integral on the angles of scattering in double series. Later in 2012, Luppé, Conoir and Norris[16] extended the Linton and Martin model to solid or porous matrix taking into account the phenomenon of conversion.

The material under consideration in this work is formed of a poro-elastic matrix in which are randomly distributed identical spherical cavities of radius $a$ greater than the pores radii. The spherical cavities are filled with a fluid identical to the fluid saturating the pores of the porous matrix. The present study proposes extension of the models of approximations of the coherent wave numbers of Waterman and Truell (WT) [6] and of Luppé, Conoir and Norris (LCN) [16] to spherical inclusions in porous media. The formula of WT is still valid in a poro-elastic medium if one neglects the phenomenon of conversions [17].

The paper is organized as follows. In the first section we study the scattering of a wave (1), (2) or (t) by a spherical cavity of radius $a$ in a poro-elastic medium in order to derive an analytical expression of the scattering coefficients in the Rayleigh limit. In the second section we derive analytical expressions of the effective properties of the coherent waves using the models of WT and LCN.

## 2  Determination of the scattering coefficients

### 2.1  The equations of the mathematical problem

We consider a spherical cavity of radius $a$ placed in a porous medium of infinite extension and filled with the same fluid as the fluid saturating the porous medium. The fluid has the density $\rho_0$ and the speed of sound $c_0$. The porous medium has the density $\rho = (1 - \psi)\rho_s + \phi\rho_0$, $\rho_s$ being the density of the solid part and $\psi$ the porosity [3]. A complete description of the porous medium according to Biot's theory includes other parameters such as tortuosity, permeability [3],[12]. Their whole set will be given in Table I for the material used for the simulation. The porous medium supports two longitudinal waves (fast and slow) of wave numbers $k_j$, $j = 1,2$ respectively and a transverse one of wave number $k_t$. The wave number in the fluid is denoted by $k_f$. When a plane harmonic wave of type $\alpha \in \{1,2,t\}$ is incident on the spherical cavity, it gives rise to scattered spherical waves of type $\beta \in \{1,2,t\}$ [9].

The physical space is related to the orthonormal reference $(O, \vec{x}, \vec{y}, \vec{z})$. The cavity is modelled by sphere of center $O$ and the incident plane wave is in the direction of $\vec{z}$. Due to the geometry of the problem we express the spatial dependence of the harmonic potentials in the spherical coordinates $r$, $\theta$, $\varphi$ the associated local reference being $(O, \vec{e}_r, \vec{e}_\theta, \vec{e}_\varphi)$. We assume a $e^{-i\omega\tau}$ dependence in the time with $\omega$ the angular frequency and $\tau$ the time. There is no dependence on $\varphi$ when the fast or slow longitudinal wave is incident but the symmetry is lost when the transverse wave is incident and the potentials depend on the variable $\varphi$[16]. The scattering of incident waves of wave numbers $k_1 = \frac{\omega}{c_1}$, $k_2 = \frac{\omega}{c_2}$ and $k_t = \frac{\omega}{c_t}$, where $c_i$, $i = 1,2,t$ are the phase velocities, is described by scattering coefficients $T_n^{\alpha\beta}$, $\alpha,\beta \in \{1,2,t\}$, $n \in \mathbb{Z}$ is the mode [16]. The displacement vector field $\vec{U}$ in the solid matrix and the relative displacement vector field $\vec{W}$ of the fluid with regard the solid are given by:

$$\vec{U} = -\sum_{\alpha=1}^{2} \vec{\nabla}\Phi_\alpha + \vec{\nabla} \wedge \overline{\Psi_t},$$
$$\vec{w} = -\sum_{\alpha=1}^{2} \gamma_\alpha \vec{\nabla}\Phi_\alpha + \gamma_t \vec{\nabla} \wedge \overline{\Psi_t}, \tag{1}$$
$$\overline{\Psi_t} = \vec{\nabla} \wedge (r\psi_t \vec{e}_r) + \vec{\nabla} \wedge \left(\vec{\nabla} \wedge (r\chi_t \vec{e}_r)\right)$$

The scalar potentials $\Phi_\alpha$, $\alpha = 1, 2$ are associated with the longitudinal waves, the scalar potentials $\psi_t$ and $\chi_t$ are associated with the transverse wave [17]. The expressions of the potentials are given in Appendix 5. The constants $\gamma_\alpha$, $\alpha = 1, 2$ are the compatibility coefficients [19]. The scattering coefficients are the coefficients of the decomposition of these potentials in series of products of Legendre polynomials with spherical Bessel or Hankel functions of the first kind.

### 2.2  The boundary conditions

When the wavelengths of of both longitudinal and transverse plane waves are much greater then the pores radii, they can propagate freely each at its characteristic velocity in the infinite poro-elastic medium. This is no longer possible if the poro-elastic medium contains obstacles of characteristic length comparable to the waves wavelengths. The phenomena of multiple scattering and mode conversion introduce differences in the behaviour of the displacement and pressure fields in the porous matrix and in the cavities.

The spherical cavities are filled with the fluid saturating the porous medium the boundary conditions are thus different depending on whether the interface between the cavity and the poro-elastic matrix is permeable or not. This leads to the boundary conditions of open pores and to those of sealed pores. Since the fluid saturating the medium is inviscid these boundary conditions are written at the interface $r = a$ for a mode $n \in \mathbb{N}$ in the form [19] :

$$
\begin{array}{rcl}
(\vec{U}_\alpha + \vec{W}_\alpha).\vec{e}_r + (\vec{U}_{sc} + \vec{W}_{sc}).\vec{e}_r & = & \vec{u}_0.\vec{e}_r \\
p_\alpha + p_{sc} & = & p_0 \\
\sigma_{\alpha,rr} + \sigma_{sc,rr} & = & -p_0 \\
\sigma_{\alpha,r\theta} + \sigma_{sc,r\theta} & = & 0
\end{array}
\tag{2}
$$

for the open pore boundary conditions and

$$
\begin{array}{rcl}
(\vec{U}_\alpha + \vec{W}_\alpha).\vec{e}_r + (\vec{U}_{sc} + \vec{W}_{sc}).\vec{e}_r & = & \vec{u}_0.\vec{e}_r \\
\vec{W}_\alpha.\vec{e}_r + \vec{W}_{sc}.\vec{e}_r & = & 0 \\
\sigma_{\alpha,rr} + \sigma_{sc,rr} & = & -p_0 \\
\sigma_{\alpha,r\theta} + \sigma_{sc,r\theta} & = & 0
\end{array}
\tag{3}
$$

for the sealed pore boundary conditions. where $\vec{u}_0$ is the displacement vector of the fluid in the spherical cavity, $p_0$ is the fluid's pressure, $\sigma_{rr}$ and $\sigma_{r\theta}$ are the components of the stress tensor at a point in the porous medium. $\alpha = 1, 2, t$ denotes the type of the incident wave and $sc$ denotes that of the scattered ones.

## 2.3 The scattering coefficients

Using the boundary conditions (3) we obtain a linear system of equations to determine the complex unknowns $T_n^{\alpha\beta}$ :

$$
M_n.\vec{X}_n^\alpha = \vec{S}_n^\alpha
\tag{4}
$$

The matrix of order 4 $M_n$ depends on the spherical Hankel and Bessel functions of the first kind, the $\vec{X}_n^\alpha$ is the unknown vector containing the scattering coefficients and the modal coefficient generated in the inclusion, $\vec{S}_n^\alpha$ is the source column vector. They are given in Appendix A. The expressions of the scattering coefficients for $n = 0,1,2$ are:

• For the fast longitudinal incident wave (1)

$$
T_0^{11} = \frac{i}{3}x_1^3[B_0^{11} - 1], \quad B_0^{11} = \frac{3(1-\Gamma_2)\rho_1 c_1^2}{(\Gamma_1-\Gamma_2)(4\rho_t c_t^2 + 3\rho_0 c_0^2)}
\tag{5}
$$

$$
T_0^{12} = \frac{i}{4}x_1^2 x_2 B_0^{12}, \quad B_0^{12} = \frac{3(\Gamma_1-1)\rho_1 c_1^2}{(\Gamma_1-\Gamma_2)(4\rho_t c_t^2 + 3\rho_0 c_0^2)}
\tag{6}
$$

$$
T_0^{1t} = 0
\tag{7}
$$

$$
T_1^{11} = \frac{i}{4}x_1^3 B_1^{11}, \quad B_1^{11} = -\frac{2(1-\Gamma_1)+2\rho_{1t}(\Gamma_t-\Gamma_2)+\rho_{2t}(1-\Gamma_1)+2\rho_{0t}(\Gamma_2-\Gamma_t)}{4(\Gamma_1-\Gamma_2)+2\rho_{1t}(\Gamma_t-\Gamma_2)+2\rho_{2t}(\Gamma_1-\Gamma_t)}
\tag{8}
$$

$$
T_1^{12} = \frac{i}{4}x_1 x_2^2 B_1^{12}, \quad B_1^{12} = -\frac{2(-1+\Gamma_1)+\rho_{1t}(3\Gamma_1-2\Gamma_t-1)+2\rho_{0t}(\Gamma_t-\Gamma_1)}{4(\Gamma_1-\Gamma_2)+2\rho_{1t}(\Gamma_t-\Gamma_2)+2\rho_{2t}(\Gamma_1-\Gamma_t)},
\tag{9}
$$

$$T_1^{1t} = \frac{i}{3} x_1 x_t^2 B_1^{1t}, \quad B_1^{1t} - \frac{\rho_{1t}(1-3\Gamma_1+2\Gamma_2)+\rho_{2t}(\Gamma_1-1)+2\rho_{0t}(\Gamma_1-\Gamma_2)}{4(\Gamma_1-\Gamma_2)+2\rho_{1t}(\Gamma_t-\Gamma_2)+2\rho_{2t}(\Gamma_1-\Gamma_t)} \tag{10}$$

$$T_2^{11} = \frac{4i}{3} x_1^3 B_2^{11}, \quad B_2^{11} = -\frac{(\Gamma_2-\Gamma_t)x_1^2}{\Delta} \tag{11}$$

$$T_2^{12} = \frac{4i}{3} x_1^2 x_2 B_2^{12}, \quad B_2^{12} = -\frac{(\Gamma_t-\Gamma_1)x_2^2}{\Delta} \tag{12}$$

$$T_2^{1t} = \frac{2i}{3} x_1^2 x_t B_2^{1t}, \quad B_2^{1t} = -\frac{(\Gamma_1-\Gamma_2)x_t^2}{\Delta} \tag{13}$$

with $\Delta = 4x_1^2(\Gamma_2 - \Gamma_t) + 4x_2^2(\Gamma_t - \Gamma_1) + x_t^2[3(\Gamma_1 - \Gamma_2) + 6(\Gamma_t \rho_{1t} + \Gamma_1 \rho_{2t} - \Gamma_t \rho_{2t} - \Gamma_2 \rho_{1t})]$.

• For the slow longitudinal incident wave (2)

$$T_0^{21} = \frac{i}{3} x_2^2 x_1 B_0^{21}, \quad B_0^{21} = -\frac{3(\Gamma_2-1)\rho_2 c_2^2}{(\Gamma_1-\Gamma_2)(4\rho_t c_t^2+3\rho_0 c_0^2)}, \tag{14}$$

$$T_0^{22} = \frac{i}{3} x_2^3 [1 - B_0^{22}], \quad B_0^{22} = \frac{3(\Gamma_1-1)\rho_2 c_2^2}{(\Gamma_1-\Gamma_2)(4\rho_t c_t^2+3\rho_0 c_0^2)}, \tag{15}$$

$$T_0^{2t} = 0, \tag{16}$$

$$T_1^{21} = -\frac{i}{3} x_2 x_1^2 B_1^{21}, \quad B_1^{21} = -\frac{2(1-\Gamma_2)+\rho_{2t}(1-3\Gamma_2+2\Gamma_t)+2\rho_{0t}(\Gamma_2-\Gamma_t)}{4(\Gamma_1-\Gamma_2)+2\rho_{1t}(\Gamma_t-\Gamma_2)+2\rho_{2t}(\Gamma_1-\Gamma_t)}, \tag{17}$$

$$T_1^{22} = \frac{i}{3} x_2^3 B_1^{22}, \quad B_1^{22} = -\frac{2(-1+\Gamma_2)+\rho_{1t}(\Gamma_2-1)+2\rho_{2t}(\Gamma_1-\Gamma_t)+2\rho_{0t}(\Gamma_t-\Gamma_1)}{4(\Gamma_1-\Gamma_2)+2\rho_{1t}(\Gamma_t-\Gamma_2)+2\rho_{2t}(\Gamma_1-\Gamma_t)}, \tag{18}$$

$$T_1^{2t} = \frac{i}{3} x_2 x_t^2 B_1^{2t}, \quad B_1^{2t} = -\frac{\rho_{2t}(-1+3\Gamma_2-2\Gamma_1)+\rho_{1t}(-\Gamma_2+1)+2\rho_{0t}(-\Gamma_2+\Gamma_1)}{4(\Gamma_1-\Gamma_2)+2\rho_{1t}(\Gamma_t-\Gamma_2)+2\rho_{2t}(\Gamma_1-\Gamma_t)}. \tag{19}$$

$$T_2^{21} = \frac{4i}{3} x_2^2 x_1 B_2^{21}, \quad B_2^{21} = -\frac{(\Gamma_t-\Gamma_2)x_1^2}{\Delta} \tag{20}$$

$$T_2^{22} = \frac{4i}{3} x_2^3 B_2^{22}, \quad B_2^{22} = -\frac{(\Gamma_t-\Gamma_1)x_2^2}{\Delta} \tag{21}$$

$$T_2^{2t} = \frac{2i}{3} x_2^2 x_t B_2^{2t}, \quad B_2^{2t} = -\frac{(\Gamma_1-\Gamma_2)x_t^2}{\Delta} \tag{22}$$

• For the transverse incident wave (t)

$$T_0^{t1} = T_0^{t2} = T_0^{tt} = 0, \tag{23}$$

$$T_1^{t1} = \frac{i}{3}x_t x_1^2 B_1^{t1}, \quad B_1^{t1} = -\frac{4(\Gamma_2-1)+2\rho_{2t}(\Gamma_t-1)+4\rho_{0t}(\Gamma_t-\Gamma_2)}{4(\Gamma_1-\Gamma_2)+2\rho_{1t}(\Gamma_t-\Gamma_2)+2\rho_{2t}(\Gamma_1-\Gamma_t)} \tag{24}$$

$$T_1^{t2} = \frac{i}{3}x_t x_2^2 B_1^{t2}, \quad B_1^{t2} = -\frac{4(1-\Gamma_1)+2\rho_{1t}(1-\Gamma_t)+4\rho_{0t}(\Gamma_1-\Gamma_t)}{4(\Gamma_1-\Gamma_2)+2\rho_{1t}(\Gamma_t-\Gamma_2)+2\rho_{2t}(\Gamma_1-\Gamma_t)} \tag{25}$$

$$T_1^{tt} = \frac{i}{3}x_t^3 B_1^{tt}, \quad B_1^{tt} = -\frac{4(\Gamma_1-\Gamma_2)(1-\rho_{0t})+2(1-\Gamma_t)(\rho_{2t}-\rho_{1t})}{4(\Gamma_1-\Gamma_2)+2\rho_{1t}(\Gamma_t-\Gamma_2)+2\rho_{2t}(\Gamma_1-\Gamma_t)} \tag{26}$$

$$T_2^{t1} = \frac{4i}{3}x_1 x_t^2 B_2^{t1}, \quad B_1^{t1} = -\frac{3(\Gamma_t-\Gamma_2)x_1^2}{\Delta} \tag{27}$$

$$T_2^{t2} = \frac{4i}{3}x_2 x_t^2 B_2^{t2}, \quad B_2^{t2} = -\frac{3(\Gamma_1-\Gamma_t)x_2^2}{\Delta} \tag{28}$$

$$T_2^{tt} = \frac{2i}{3}x_t^3 B_2^{tt}, \quad B_2^{tt} = -\frac{3(\Gamma_2-\Gamma_1)x_t^2}{\Delta} \tag{29}$$

with $\Gamma_\alpha = 1 + \gamma_\alpha$, $\rho_{\alpha t} = \frac{\rho_\alpha}{\rho_t}$, $\alpha = 1, 2, t$.

$\gamma_\alpha$ and $\rho_\alpha$, $\alpha = 1, 2, t$ are respectively the compatibility coefficients and the acoustic densities [11][19]. Letting $x_\alpha = k_\alpha a$, $\epsilon = 1, 2, 4$ we have, in the Rayleigh limit, $x_\alpha \ll 1$ and we can compute approximations of the scattering coefficients using the Taylor series developments in the neighbourhood of zero. We get:

$$\begin{cases} T_n^{\alpha\beta} = \frac{i\epsilon}{3}x_\alpha^2 x_\beta B_n^{\alpha\beta} + O(x_\alpha^5) \quad ou \quad T_n^{\alpha\beta} = \frac{i\epsilon}{3}x_\alpha x_\beta^2 B_n^{\alpha\beta} + O(x_\alpha^5) \quad n = 0, 1, 2 \\ T_n^{\alpha\beta} = O(x_\alpha^3), \quad t_n^{tt} = O(x_t^5) \quad n \geq 3 \end{cases} \tag{30}$$

We find that in the quasi-static limit, $\frac{\omega}{\omega_c} << 1$ ($\omega_c = \frac{\eta}{\rho_0 \kappa}$ is the characteristic angular frequency, $\kappa$ the permeability and $\eta$ the kinematic viscosity of fluid pore ), when the porosity tends towards to zero, the scattering coefficients tend towards the scattering coefficients found by C.F. Ying, R. Truell [4] GC Gaunaurd and H. Ãœberall [10], Lepert [17] (by removing the elasticity of the spherical cavity) then by Norman G. Einspruch and Rohn Truell [5] (for longitudinal incident waves).

To check the validity of the approximate values of the $T_n^{\alpha\beta}$, we perform a numerical simulation of the scattering coefficients $T_0^{11}$, $T_1^{11}$, $T_2^{11}$ and $T_2^{1t}$ versus the frequency. We take the QF20 [19] as a porous medium which obeys the theory of Biot [3] and the spherical cavity has a radius $a = 10^{-4}m$. At sufficiently low frequencies, the continuous curves (exact coefficients) and the dotted curves (approximations) differ little (Figure 1). Therefore the approximate scattering coefficients are sufficiently precise in the low frequency (Rayleigh limit) domain to be used to determine the effective properties of the medium.

| | |
|---|---|
| Bulk modulus of grains | $36,6.10^9 Pa$ |
| Dried frame bulk modulus | $9,47.10^9 Pa$ |
| Dried frame shear modulus | $7,63.10^9 Pa$ |
| Solid density | $2760 kg.m^{-3}$ |
| Bulk modulus of water | $2,22.10^9 Pa$ |
| Density of water | $1000 kg.m^{-3}$ |
| Viscosity of water | $1,14.10^{-3} kg.m^{-1}.s^{-1}$ |
| Porosity | $0,402$ |
| Permeability | $1,68.10^{-11} m^2$ |
| Pore radius | $3,26.10^{-5} m$ |
| Tortuosity | $1,89$ |

Table 1: Values of the parameters of QF20 and water

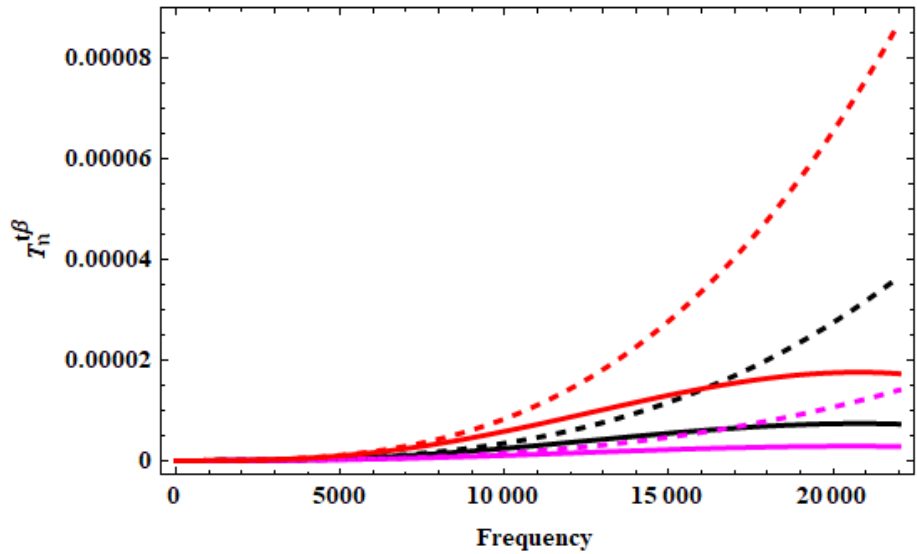

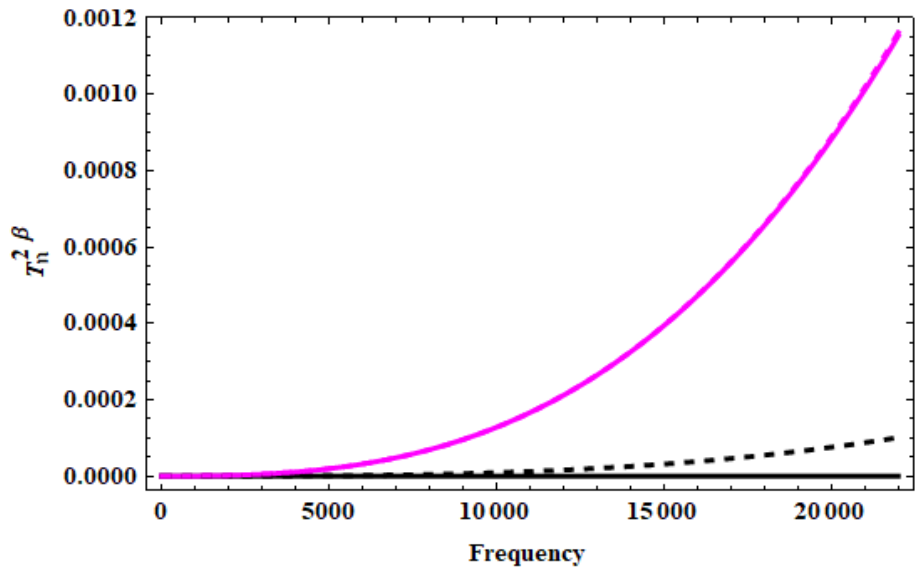

Figure 1: Comparison of exact (continuous curves) and approximated (dotted curves) scattering coefficients for the incident slow wave in mode $n = 0$ (bottom figure) and transverse wave in mode $n = 2$ (top figure) versus frequency (Hz), $\beta = 1$ for fast wave (black ), $\beta = 2$ for slow wave (magenta), $\beta = t$ for transverse wave (red).

## 3   Effective properties

We use two multiple scattering models to compute the effective properties of the coherent waves namely the Waterman-Truell model (WT) and the the Luppé-Conoir-Norris model (LCN).

### 3.1   The model of Waterman and Truell (WT)

The effective wavenumber of the WT model applied to a porous matrix for each wave type (compressional fast, compressional slow and transverse) is written as[17]

$$\left(\frac{\xi_\alpha}{k_\alpha}\right)^2_{WT} = 1 + \frac{4\pi}{k_\alpha^2} n_0 f_{\alpha\alpha}(0) + \frac{4\pi^2}{k_\alpha^4} n_0^2 (f_{\alpha\alpha}(0)^2 - f_{\alpha\alpha}(\pi)^2). \tag{31}$$

where $\alpha = 1, 2, t$. The far-field scattered amplitudes $f_{\alpha\alpha}(0)$ et $f_{\alpha\alpha}(\pi)$ describe respectively the forward scattering and backward scattering of a sphere:

$$f_{\alpha\alpha}(0) = \frac{1}{ik_\alpha} \sum_{n=0}^{+\infty} (2n+1) T_n^{\alpha\alpha}, \quad f_{\alpha\alpha}(\pi) = \frac{1}{ik_\alpha} \sum_{n=0}^{+\infty} (-1)^n (2n+1) T_n^{\alpha\alpha}, \quad \alpha = 1, 2 \tag{32}$$

$$f_{tt}(0) = \frac{1}{ik_t} \sum_{n=1}^{+\infty} \frac{2n+1}{2} (T_n^{tt} + t_n^{tt}), \quad f_{tt}(\pi) = \frac{1}{ik_t} \sum_{n=1}^{+\infty} (-1)^n \left(\frac{2n+1}{2}\right) (T_n^{tt} + t_n^{tt}) \tag{33}$$

$t_n^{tt}$ being the modal coefficient associated with the scalar potential $\chi_t$ (appendix 6). By replacing the scattering coefficients by their expressions((5)- (28)) in the case $\alpha = 1, t$, we get

$$\left(\frac{\xi_1}{k_1}\right)^2_{WT} = 1 + \phi(B_0^{11} - 1 + 3B_1^{11} + 20B_2^{11}) + \phi^2(3(B_0^{11} - 1)B_1^{11} + 60B_1^{11}B_2^{11}) \tag{34}$$

$$\left(\frac{\xi_t}{k_t}\right)^2_{WT} = 1 + \phi\left(\frac{3}{2}B_1^{tt} + 5B_2^{tt}\right) + \frac{15}{2}\phi^2 B_1^{tt} B_2^{tt} \tag{35}$$

where $\phi = \frac{4\pi a^3 n_0}{3}$, $n_0$ is the concentration per unit volume and $t_n^{tt} = O(k_t^5 a^5) \ \forall n \geq 1$ (85). We deduce the case of the slow longitudinal wave by replacing the fast longitudinal wave (1) by the slow longitudinal one (2) in the equation (34). We can rewrite the equations (34) and (35) in the form

$$\left(\frac{\xi_1}{k_1}\right)^2_{WT} = (1 + 3\phi B_1^{11})\left(1 + \phi(B_0^{11} - 1 + 20B_2^{11})\right) \tag{36}$$

$$\left(\frac{\xi_t}{k_t}\right)^2_{WT} = \left(1 + \frac{3}{2}\phi B_1^{tt}\right)(1 + 5\phi B_2^{tt}). \tag{37}$$

Hence the effective mass densities and the dynamic bulk, shear moduli for the fast($\alpha = 1$) and shear($\alpha = t$) coherent waves are given by:

$$\left(\frac{\rho_{1,eff}^{WT}}{\rho_1}\right) = 1 + 3\phi B_1^{11} \tag{38}$$

$$\left(\frac{\rho_{2,eff}^{WT}}{\rho_2}\right) = 1 + 3\phi B_1^{22} \tag{39}$$

$$\left(\frac{\rho_{t,eff}^{WT}}{\rho_t}\right) = 1 + \frac{3}{2}\phi B_1^{tt} \tag{40}$$

$$\left(\frac{M_1}{M_{1,eff}^{WT}}\right) = 1 + \phi(B_0^{11} - 1 + 20B_2^{11}). \tag{41}$$

$$\left(\frac{M_2}{M_{2,eff}^{WT}}\right) = 1 + \phi(B_0^{22} - 1 + 20B_2^{22}). \tag{42}$$

$$\left(\frac{M_t}{M_{t,eff}^{WT}}\right) = 1 + 5\phi B_2^{tt} \tag{43}$$

where $M_1$, $M_2$ are the bulk moduli associated respectively with the fast longitudinal wave (1) and the slow longitudinal wave (2), $M_t$ is the shear modulus of the medium. $M_1$ and $M_t$ do not depend on the frequency[19]. The effective mass density $\rho_{\alpha,eff}^{WT}$ and the effective bulk or shear modulus $M_{\alpha,eff}^{WT}$, $\alpha = 1$, $2$, $t$, are such that $\left(\frac{\xi_\alpha}{k_\alpha}\right)^2 = \frac{\rho_{\alpha,eff}^{WT}}{\rho_\alpha} \times \frac{M_\alpha}{M_{\alpha,eff}^{WT}}$.

The quasi-static limits of the coefficients $B_n^{\alpha\beta}$ are

$$B_0^{11} \rightarrow \frac{3H}{3K_0 + 4\mu}, \quad B_0^{12} \rightarrow 0,$$

$$B_1^{11} \rightarrow \frac{\rho_0 - \rho}{3\rho}, \quad B_1^{tt} \rightarrow \frac{2(\rho_0 - \rho)}{3\rho}, B_1^{12} \rightarrow 0$$

$$B_2^{11} \rightarrow \frac{\mu}{9H - 4\mu}, \quad B_2^{tt} \rightarrow \frac{3H}{9H - 4\mu}, \quad B_2^{12} \rightarrow 0,$$

$$B_1^{1t} \rightarrow -\frac{\rho_0 - \rho}{3\rho}, B_1^{t1} \rightarrow -\frac{2(\rho_0 - \rho)}{3\rho}, B_1^{t2} \rightarrow 0,$$

$$B_2^{1t} \rightarrow \frac{H}{4\mu - 9H}, \quad B_2^{t1} \rightarrow \frac{3\mu}{4\mu - 9H}, \quad B_2^{t2} \rightarrow 0,$$

$$B_0^{22} \rightarrow 0, \quad B_2^{22} \rightarrow 0, \quad B_0^{t2} \rightarrow 0, \quad B_1^{22} \rightarrow \frac{1}{2}$$

with $K_0 = \rho_0 c_0^2$, $H$ is static bulk modulus and $\mu$ the static shear modulus of the poro-elastic medium. Putting the quasi-static coefficients in the equations (38) to (43) and using the approximation $M_1 \simeq H$, $M_t \simeq \mu$[9], [19] yields

$$\left(\frac{\rho_{1,eff}^{WT}}{\rho_1}\right)_{\omega \rightarrow 0} = 1 + \phi\Delta\rho \tag{44}$$

$$\left(\frac{\rho_{t,eff}^{WT}}{\rho_t}\right)_{\omega \rightarrow 0} = 1 + \phi\Delta\rho \tag{45}$$

$$\left(\frac{H}{M_{1,eff}^{WT}}\right)_{\omega\to0} = 1 + \phi\left(\frac{3H-4\mu-3K_0}{4\mu+3K_0} + \frac{20\mu}{9H-4\mu}\right) \qquad (46)$$

$$\left(\frac{\mu}{M_{t,eff}^{WT}}\right)_{\omega\to0} = 1 + \phi\left(\frac{15H}{9H-4\mu}\right) \qquad (47)$$

where $\Delta\rho = (\rho_0 - \rho)/\rho$. $H$ and $\mu$ depend on porosity.

For an elastic medium of Lamé coefficients $\lambda_s$, $\mu_s$, let $K_s = \lambda_s + 2\frac{\mu_s}{3}$ be the bulk modulus and $\rho_s$ the mass density . Then the phase velocities of the longitudinal and shear waves are respectively $c_L = \sqrt{\frac{(\lambda_s+2\mu_s)}{\rho_s}}$ and $c_T = \sqrt{\frac{\mu_s}{\rho_s}}$.

When the porosity tends towards zero, we have $\mu \to \mu_s$, $\rho \to \rho_s$ and $H \to K_s + 4\frac{\mu_s}{3}$. Hence the effective properties derived from the quasi-static equations when the porosity tends towards infinity are:

$$\frac{\rho_L^{WT}}{\rho_L} = \frac{\rho_T^{WT}}{\rho_T} = 1 + \phi\Delta\rho_s \qquad (48)$$

$$\frac{1}{M_{L,eff}^{WT}} = \frac{1}{\lambda_s+2\mu_s} + \frac{\phi}{\lambda_s+2\mu_s}\left(\frac{3\lambda_s+2\mu_s-3K_0}{4\mu+3K_0} + \frac{20\mu_s}{9\lambda_s+14\mu_s}\right) \qquad (49)$$

$$\frac{1}{M_{T,eff}^{WT}} = \frac{1}{\mu_s} + \frac{15\phi}{\mu_s}\frac{\lambda_s+2\mu_s}{9\lambda_s+14\mu_s} \qquad (50)$$

exactly the same obtained by G. Lepert[17] for the elastic medium of of Lamé coefficients $\lambda_s$ and $\mu_s$.

Curves of the effective phase velocities $c_{\alpha,eff} = Re\left(\frac{\omega}{\xi_\alpha}\right)$ ($\alpha = 1,\ 2,\ t$) are presented in figure (Figure 2) and (Figure 3). The radius of the spherical cavities is $10^{-3}m$ and the porous medium used is QF20.

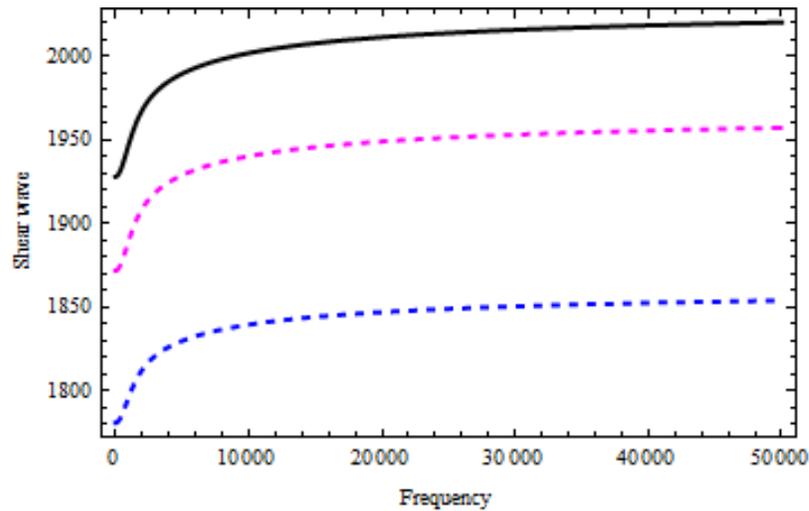

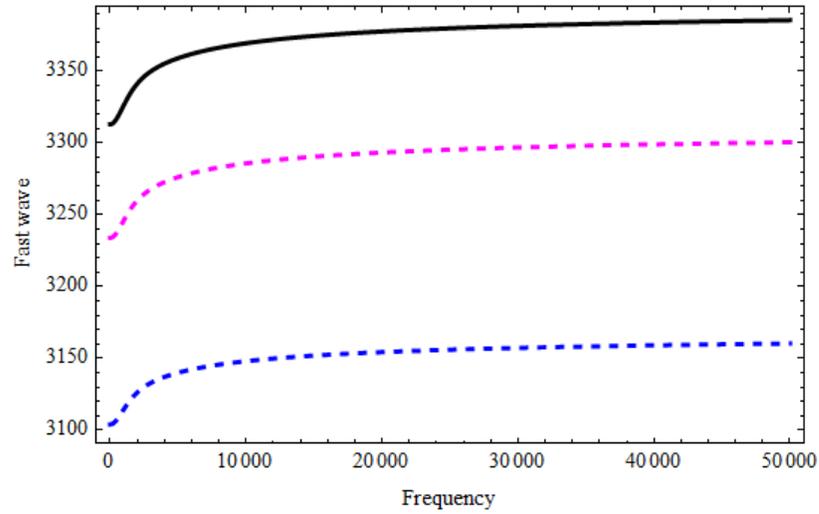

Figure 2: Curves of the phase velocities of the transverse wave and fast wave versus frequency(Hz) for concentrations $\phi \simeq 12\%$ (blue curve) and $\phi \simeq 4\%$ (magenta curve), the black curve corresponds to a poro-elastic medium without obstacles.

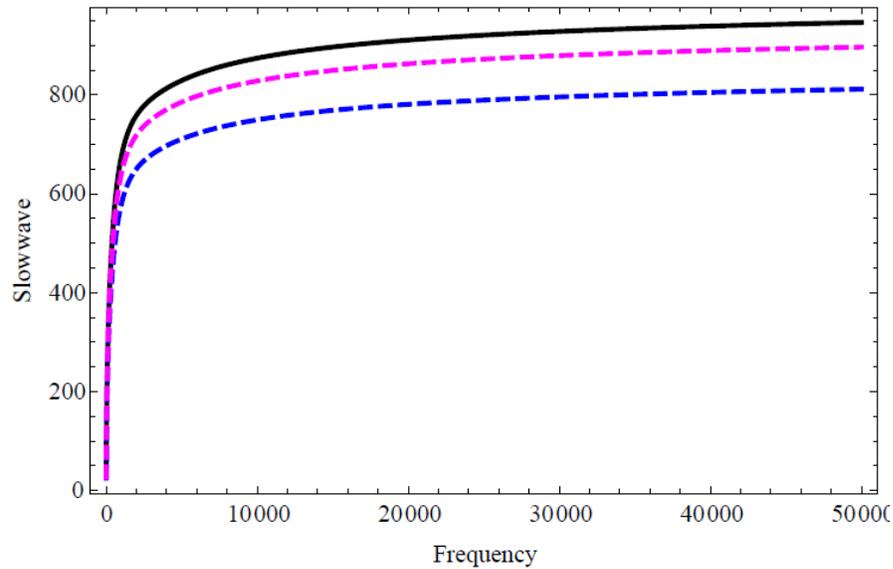

Figure 3: Curves of the phase velocity of the slow wave versus frequency(Hz) for concentrations $\phi \simeq 12\%$ (blue curve) and $\phi \simeq 4\%$ (magenta curve), the black curve corresponds to a poro-elastic medium without obstacles.

## 3.2 The model of Luppé, Conoir and Norris (LCN)

The model of Luppé, Conoir and Norris leads to the following formula of the effective wave number for each kind of wave $\alpha = 1,\ 2,\ t$:

$$\xi_\alpha^2 = k_\alpha^2 + n_0 \delta_1^\alpha + n_0^2 \left( \delta_2 + \delta_2^{\alpha(c)} \right), \tag{51}$$

$$\delta_1^\alpha = 4\pi f_{\alpha\alpha}(0), \tag{52}$$

$$\delta_2^\alpha = -\frac{1}{2} \left( \frac{4\pi}{k_\alpha} \right)^4 \sum_{n,m=0}^{+\infty} K_{nm} T_n^{\alpha\alpha} T_m^{\alpha\alpha} \tag{53}$$

$$\delta_2^{\alpha(c)} = \sum_{\beta \neq \alpha} \frac{16\pi^2}{k_\alpha k_\beta \left( k_\beta^2 - k_\alpha^2 \right)} \sum_{n=0}^{+\infty} \sum_{\nu=0}^{+\infty} \sum_{q=|n-\nu|}^{n+\nu} (2n+1)(2\nu+1) \left( \frac{k_\alpha}{k_\beta} \right)^q T_n^{\beta\alpha} T_\nu^{\alpha\beta} G(0,\nu;0,n;q), \tag{54}$$

$$K_{nm} = \left( \frac{1}{4\pi} \right)^{3/2} \sqrt{(2n+1)(2\nu+1)} \sum_{q=|n-\nu|}^{n+\nu} q\sqrt{2q+1} \, G(n,0\ ;\ \nu,0\ ;\ q) \tag{55}$$

$$G(0,n;0,\nu;q) = \frac{G(n,0\ ;\ \nu,0\ ;\ q)}{\sqrt{\frac{(2n+1)(2\nu+1)}{4\pi(2q+1)}}} \tag{56}$$

where $G$ is the Gaunt coefficient. The sum on the index $q$ is in steps of two with $n+m+q$ even. Terms $\delta_1^\alpha$ and $\delta_2$ only use scattering coefficients $T_n^{\alpha\alpha}$. The mode conversions represented by the scattering coefficients $T_n^{\alpha\beta}$ occurring during each scattering process are taken into account in the term $\delta_2^{\alpha(c)}$. As previously stated $\alpha$ represents the incident wave and $\beta$ the scattered one.

Developments at low frequencies of the scattering coefficients show that the scattering coefficient $T_n^{\alpha\beta}$ are of order $(k_\alpha^3 a^3)$ if $n \in \{0,1,2\}$ and of $(k_\alpha^5 a^5)$ if $n \geq 3$. Consequently, the series in the equations representing the effective wave numbers, can be replaced by partial sums from 0 to 2. We thus have:

$$\delta_1^\alpha = \frac{4\pi}{ik_\alpha} [T_0^{\alpha\alpha} + 3T_1^{\alpha\alpha} + 5T_2^{\alpha\alpha}], \tag{57}$$

$$\delta_2^\alpha = -\frac{16\pi^2}{2k_\alpha^4} \left[ 6T_0^{\alpha\alpha} T_1^{\alpha\alpha} + 20T_0^{\alpha\alpha} T_2^{\alpha\alpha} + 12T_1^{\alpha\alpha} T_1^{\alpha\alpha} + 66T_1^{\alpha\alpha} T_2^{\alpha\alpha} + \frac{460}{7} T_2^{\alpha\alpha} T_2^{\alpha\alpha} \right], \tag{58}$$

$$\delta_2^{\alpha(c)} = -\frac{16\pi^2}{2k_\alpha^4} \sum_{\beta \neq \alpha} \frac{2k_\alpha^3}{k_\beta \left( k_\alpha^2 - k_\beta^2 \right)} \left\{ T_0^{\alpha\beta} T_0^{\beta\alpha} + \left( 3 + 6k_{\alpha\beta}^2 \right) T_1^{\beta\alpha} T_1^{\alpha\beta} + \left( 5 + \frac{50}{7} k_{\alpha\beta}^2 + \frac{90}{7} k_{\alpha\beta}^4 \right) T_2^{\beta\alpha} T_2^{\alpha\beta} \right.$$
$$\left. + 3k_{\alpha\beta} \left( T_0^{\beta\alpha} T_1^{\alpha\beta} + T_1^{\beta\alpha} T_0^{\alpha\beta} \right) + 5k_{\alpha\beta}^2 \left( T_2^{\beta\alpha} T_0^{\alpha\beta} + T_0^{\beta\alpha} T_2^{\alpha\beta} \right) + \left( 6k_{\alpha\beta} + 9k_{\alpha\beta}^3 \right) \left( T_2^{\beta\alpha} T_1^{\alpha\beta} + T_1^{\beta\alpha} T_2^{\alpha\beta} \right) \right\} \tag{59}$$

$$\delta_1^t = \frac{4\pi}{ik_t} \left[ \frac{3}{2} T_1^{tt} + \frac{5}{2} T_2^{tt} \right], \tag{60}$$

$$\delta_2^t = -\frac{16\pi^2}{2k_t^4}\Big[12T_1^{tt}T_1^{tt} + 66T_1^{tt}T_2^{tt} + \frac{460}{7}T_2^{tt}T_2^{tt}\Big], \tag{61}$$

$$\delta_2^{t(c)} = -\frac{16\pi^2}{2k_t^4}\sum_{\beta\neq t}\frac{2k_t^3}{k_\beta\left(k_t^2-k_\beta^2\right)}\Big\{\left(3+6k_{t\beta}^2\right)T_1^{\beta t}T_1^{t\beta} + \left(5+\frac{50}{7}k_{t\beta}^2 + \frac{90}{7}k_{t\beta}^4\right)T_2^{\beta t}T_2^{t\beta}$$
$$+ \left(6k_{t\beta} + 9k_{t\beta}^3\right)\left(T_2^{\beta t}T_1^{t\beta} + T_1^{\beta t}T_2^{t\beta}\right)\Big\}. \tag{62}$$

where $k_{\alpha\beta} = \frac{k_\alpha}{k_\beta}$ and $\alpha, \beta = 1, 2$. At Rayleigh limit, we obtain the following approximate expressions for the incident wave $\alpha = 1, 2$:

$$\delta_1^\alpha = \frac{\phi k_\alpha^2}{n_0}\big[B_0^{\alpha\alpha} - 1 + 3B_1^{\alpha\alpha} + 20B_2^{\alpha\alpha}\big], \tag{63}$$

$$\delta_2^\alpha = \frac{\phi^2 k_\alpha^2}{n_0^2}\Big[3(B_0^{\alpha\alpha}-1)B_1^{\alpha\alpha} + 40(B_0^{\alpha\alpha}-1)B_2^{\alpha\alpha} + 6B_1^{\alpha\alpha}B_1^{\alpha\alpha} + 132B_1^{\alpha\alpha}B_2^{\alpha\alpha} + \frac{3680}{7}B_2^{\alpha\alpha}B_2^{\alpha\alpha}\Big], \tag{64}$$

$$\delta_2^{\alpha(c)} = \frac{\phi^2 k_\alpha^2}{n_0^2}\Big\{\frac{1}{k_{\alpha\beta}^2-1}\big[B_0^{\alpha\beta}B_0^{\beta\alpha} + \left(3+6k_{\alpha\beta}^2\right)B_1^{\beta\alpha}B_1^{\alpha\beta} + 16\left(5+\frac{50}{7}k_{\alpha\beta}^2 + \frac{90}{7}k_{\alpha\beta}^4\right)B_2^{\beta\alpha}B_2^{\alpha\beta}$$
$$+3\big(B_0^{\beta\alpha}B_1^{\alpha\beta} + k_{\alpha\beta}^2 B_1^{\beta\alpha}B_0^{\alpha\beta}\big) + 20k_{\alpha\beta}^2\big(B_2^{\beta\alpha}B_0^{\alpha\beta} + B_0^{\beta\alpha}B_2^{\alpha\beta}\big) + \left(24k_{\alpha t} + 36k_{\alpha t}^3\right)\big(B_2^{\beta\alpha}B_1^{\alpha\beta} + k_{\alpha\beta}^2 B_1^{\beta\alpha}B_2^{\alpha\beta}\big)\big]$$
$$+\frac{1}{k_{\alpha t}^2-1}\big[(3+6k_{\alpha t}^2)B_1^{t\alpha}B_1^{\alpha t} + 8\left(5+\frac{50}{7}k_{\alpha t} + \frac{90}{7}k_{\alpha t}^4\right)T_2^{t\alpha}T_2^{\alpha t} + (12k_{\alpha t} + 18k_{\alpha t}^3)(2B_2^{t\alpha}B_1^{\alpha t} + k_{\alpha t}^2 B_1^{t\alpha}B_2^{\alpha t})\big]\Big\} \tag{65}$$

, for $\alpha = t$ we have

$$\delta_1^t = \frac{\phi k_t^2}{n_0}\Big[\frac{3}{2}B_1^{tt} + 5B_2^{tt}\Big], \tag{66}$$

$$\delta_2^t = \frac{\phi^2 k_t^2}{n_0^2}\Big[6B_1^{tt}B_1^{tt} + 66B_1^{tt}B_2^{tt} + \frac{920}{7}B_2^{tt}B_2^{tt}\Big], \tag{67}$$

$$\delta_2^{t(c)} = \frac{\phi^2 k_t^2}{n_0^2}\sum_{\beta\neq t}\frac{1}{k_{t\beta}^2-1}\Big\{\left(3+6k_{t\beta}^2\right)B_1^{\beta t}B_1^{t\beta} + 8\left(5+\frac{50}{7}k_{t\beta}^2 + \frac{90}{7}k_{t\beta}^4\right)B_2^{\beta t}B_2^{t\beta}$$
$$+\left(12k_{t\beta} + 18k_{t\beta}^3\right)\left(k_{t\beta}^2 B_2^{\beta t}B_1^{t\beta} + 2B_1^{\beta t}B_2^{t\beta}\right)\Big\} \tag{68}$$

where $\beta = 1, 2$.

In the quasi-static limit when the porosity tends towards zero, we find the effective wave numbers and the phase velocities of the acoustic waves in an elastic medium [18] by removing the elasticity of the spherical inclusions.

Curves of the effective phase velocities $c_{\alpha,eff} = Re\left(\frac{\omega}{\xi_\alpha}\right)$ $(\alpha = 1, 2, t)$ deduced from the effective wave numbers $\left(\frac{\xi_\alpha}{k_\alpha}\right)^2$ computed in the scope of the LCN's model are presented in figures (Figure 4) and (Figure 5). The radius of the spherical cavities is $10^{-3}m$ and the porous medium used is QF20.

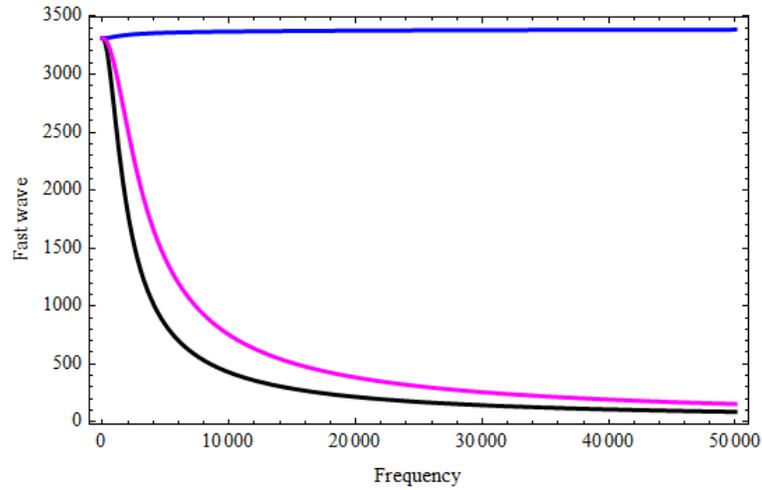

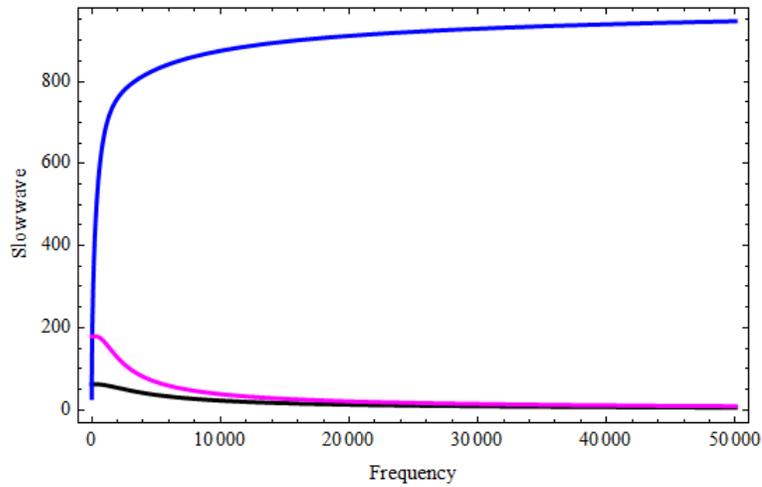

Figure 4: Curves of the phase velocities of the fast wave and the slow wave versus frequency(Hz) for concentrations $\phi \simeq 12{,}57\%$ (black curve) and $\phi \simeq 4{,}2\%$ (magenta curve), the blue curve corresponds to a poro-elastic medium without obstacles.

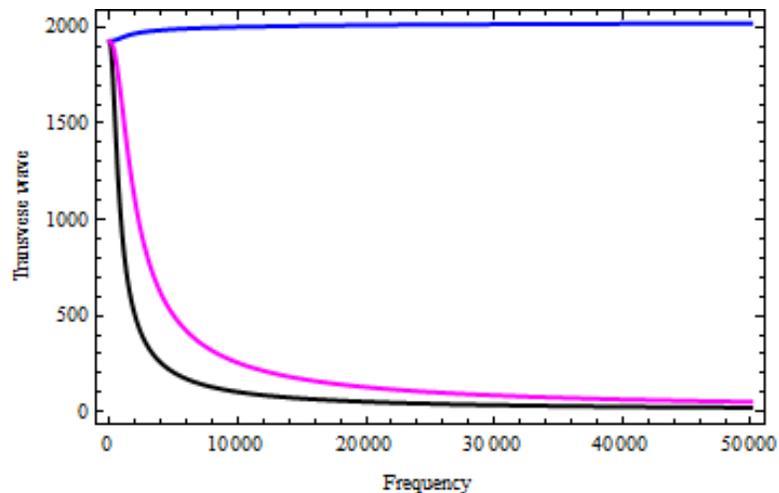

Figure 5: Curves of the phase velocity of the transverse wave versus frequency(Hz) for concentrations $\phi \simeq 12{,}57\%$ (black curve) and $\phi \simeq 4{,}2\%$ (magenta curve), the blue curve corresponds to a poro-elastic medium without obstacles.

## 4  Conclusion

Rayleigh limit approximations have been obtained for the effective wave numbers of coherent waves propagating in a material formed of a porous medium containing a random distribution of identical spherical cavities filled with the same fluid as the fluid saturating the porous matrix. Using the WT model, explicit expressions for the effective mass densities and moduli for each kind of coherent wave have been deduced. As validation result, we recover the effective properties of an elastic medium as the limit when the porosity tends towards zero of the effective properties of the material in the quasi-static limit. For the LCN's model, analytical expressions have been found for the wave numbers of the coherent waves which converge when the porosity tends to zero in the quasi-static limit to the wave number of the coherent wave for the elastic medium. However for this model the determination of the effective mass densities and moduli from the expressions of the effective wave numbers is difficult to achieve.

## Appendix

## A  Expressions of the potentials

### A.1  Longitudinal wave incident $\alpha = 1, 2$

When the fast or slow wave is incident, the potentials functions describing the incident wave can be written :

$$\Phi_\alpha(r,\theta) = \sum_{n=0}^\infty (-i)^n (2n+1) j_n(k_\alpha r) P_n(cos\theta), \quad \alpha = 1, 2 \tag{69}$$

$$\psi_t(r,\theta) = \chi_t(r,\theta) = 0. \tag{70}$$

The potentialsf of the scattered wave in the porous medium are given by

$$\Phi_{\alpha\beta}(r,\theta) = \sum_{n=0}^\infty (-i)^n (2n+1) T_n^{\alpha\beta} h_n(k_\beta r) P_n(cos\theta), \quad \alpha, \beta = 1, 2 \tag{71}$$

$$\psi_{\alpha\, t}(r,\theta) = \sum_{n=0}^\infty (-i)^n (2n+1) T_n^{\alpha t} j_n(k_t r) P_n(cos\theta), \quad \alpha = 1, 2 \tag{72}$$

$$\chi_{\alpha t}(r,\theta) = 0 \quad \alpha = 1, 2. \tag{73}$$

The potential for the wave within the sphere are given by

$$\Phi_{\alpha\, 0}(r,\theta) = \sum_{n=0}^\infty (-i)^n (2n+1) B_n^{\alpha 0} j_n(k_0 r) P_n(cos\theta), \quad \alpha = 1, 2. \tag{74}$$

$P_n$ Legendre polynomial of order n, $P_n^1$ Legendre polynomial of order n and 1.

### A.2  Transversal wave incident $t$

The potentials functions describing the incident wave are:

$$\psi_t(r,\theta,\varphi) = \sum_{n=1}^\infty (-i)^n \frac{2n+1}{n(n+1)} j_n(k_t r) P_n^1(cos\theta) cos\varphi \tag{75}$$

$$\chi_t(r,\theta,\varphi) = \sum_{n=1}^{\infty} (-i)^n \frac{2n+1}{n(n+1)} \frac{ij_n(k_tr)}{k_t} P_n^1(cos\theta)sin\varphi. \tag{76}$$

The potentialsf of the scattered wave in the porous medium are given by

$$\Phi_{t\beta}(r,\theta,\varphi) = \sum_{n=1}^{\infty} (-i)^n \frac{2n+1}{n(n+1)} T_n^{t\beta} h_n(k_\beta r) P_n^1(cos\theta)cos\varphi \quad \beta = 1,\ 2 \tag{77}$$

$$\psi_{tt}(r,\theta,\varphi) = \sum_{n=1}^{\infty} (-i)^n \frac{2n+1}{n(n+1)} T_n^{tt} h_n(k_tr) P_n^1(cos\theta)sin\varphi. \tag{78}$$

$$\chi_{tt}(r,\theta,\varphi) = \sum_{n=1}^{\infty} (-i)^n \frac{2n+1}{n(n+1)} t_n^{tt} \frac{ih_n(k_tr)}{k_t} P_n^1(cos\theta)sin\varphi. \tag{79}$$

The potential for the wave within the sphere are given by

$$\Phi_{t\,0}(r,\theta,\varphi) = \sum_{n=1}^{\infty} (-i)^n \frac{2n+1}{n(n+1)} B_n^{t\,0} j_n(k_0r) P_n^1(cos\theta)sin\varphi. \tag{80}$$

## B  Matrix equation

The matrix equation for determining the scattering coefficients is

$$M_n.\vec{X}_n^\alpha = \vec{S}_n^\alpha, \tag{81}$$

where

$$M_n =$$

$$\begin{bmatrix} -(\gamma_1+1)x_1h_n{}'(x_1) & -(\gamma_2+1)x_2h_n{}'(x_2) & n(n+1)(\gamma_t+1)h_n(x_t) & x_0j_n{}'(x_0) \\ -\gamma_1x_1h_n{}'(x_1) & -\gamma_2x_2h_n{}'(x_2) & n(n+1)\gamma_th_n(x_t) & 0 \\ 4x_1h_n{}'(x_1)+(\rho_{1t}x_t^2-2n(n+1))h_n(x_1) & 4x_2h_n{}'(x_2)+(\rho_{2t}x_t^2-2n(n+1))h_n(x_2) & 2n(n+1)[x_th_n{}'(x_t)-h_n(x_t)] & -\rho_{0t}x_t^2j_n(x_0) \\ x_1h_n{}'(x_1)-h_n(x_1) & x_2h_n{}'(x_2)-h_n(x_2) & x_th_n{}'(x_t)+(1-n(n+1)+x_t^2/2)h_n(x_t) & 0 \end{bmatrix}$$

$$\vec{X}_n^\alpha = \begin{pmatrix} T_n^{\alpha\,1} \\ T_n^{\alpha\,2} \\ T_n^{\alpha\,t} \\ A_n^{\alpha\,0} \end{pmatrix} \tag{82}$$

$$\vec{S}_n^\alpha = \begin{bmatrix} (1+\gamma_\alpha)x_\alpha j_n{}'(x_\alpha) \\ \gamma_\alpha x_\alpha j_n{}'(x_\alpha) \\ -4x_\alpha j_n{}'(x_\alpha)-(\rho_{\alpha\,t})x_t^2-2n(n+1))j_n(x_\alpha) \\ -x_\alpha j_n{}'(x_1)+j_n(x_\alpha) \end{bmatrix} \quad \alpha = 1,\ 2 \tag{83}$$

$$\vec{S}_n^t = \begin{bmatrix} -n(n+1)(1+\gamma_t)j_n(x_t) \\ -n(n+1)\gamma_tj_n(x_t) \\ -2n(n+1)(x_tj_n{}'(x_t)-j_n(x_t)) \\ -x_tj_n{}'(x_t)-(1-n(n+1)+\frac{x_t^2}{2})j_n(x_t) \end{bmatrix} \tag{84}$$

$$(x_tj_n{}'(x_t)-j_n(x_t)) + (x_th_n{}'(x_t)-h_n(x_t))t_n^{tt} = 0, \quad n \geq 1 \tag{85}$$

$\alpha = 1,\ 2,\ t$ is incident wave, $x_\alpha = k_\alpha a$, $\rho_{\alpha\,t} = \frac{\rho_\alpha}{\rho_t}$, $\gamma_\alpha$ are the compatibility coefficients[19], $t_n^{tt}$ is the modal coefficient associated with the potential $\chi$ for $n \geq 1$, $j_n$ is the spherical Bessel function of the first kind and $h_n$ is the spherical Hankel function of the first kind.